\documentclass[12pt]{article}
\usepackage{color}
\usepackage{hyperref}
\usepackage{amsmath}
\begin{document}

\begin{titlepage}

\title{Accelerated expansion of the Universe in a \\higher dimensional modified gravity with\\ Euler-Poincar\'{e} terms}

\author{ \"{O}zg\"{u}r Akarsu\footnote{akarsuo@itu.edu.tr} \footnote{Present address: Department of Physics, {\.I}stanbul Technical University, 34469, Maslak,  {\.I}stanbul, Turkey} , Tekin Dereli\footnote{tdereli@ku.edu.tr}, Neslihan Oflaz\footnote{noflaz@ku.edu.tr}
\\ {\small Department of Physics, Ko\c{c} University, 34450 Sar{\i}yer, \.{I}stanbul, Turkey }}

\date{ }

\maketitle

\begin{abstract}

\noindent A higher dimensional modified gravity theory with an action that includes dimensionally continued Euler-Poincar\'{e} forms
up to second order in curvatures is considered. The variational field equations are derived. Matter in the universe at large scales is modeled by a fluid satisfying an equation of state with dimensional dichotomy.
We study solutions that describe higher dimensional steady state cosmologies with  constant volume for which the three dimensional external space is expanding at an accelerated rate while the (compact) internal space is contracting. We showed that the second order Euler-Poincar\'{e} term in the constructions of higher dimensional steady state cosmologies could be crucial.

\end{abstract}

\vskip 2cm

\thispagestyle{empty} 
\end{titlepage}

\section{Introduction}

The idea of considering an early inflationary epoch during the evolution of the Universe that is characterized by an  exponential expansion (that lasted till $\sim 10^{-35}$ seconds after the Big Bang, corresponding to energy scales $\sim 10^{16}\,{\rm GeV}$) is successful  in resolving the problems of standard Big Bang cosmology such as the horizon and flatness problems. Furthermore the same idea provides strong clues for an explanation of the origin of large scale temperature fluctuations observed in the cosmic microwave background radiation \cite{Starobinsky80,Guth80,Linde82,Albrect82}. The actual models of inflation have several variants that are mostly based on general relativity in which the inflation would be driven by scalar field(s) with ad hoc potential(s); see for instance Ref.\cite{Encyclopaedia} for a recent review. We do not yet have a concrete and unique realization of inflation from a fundamental theory. Leaving aside the accelerated expansion in the early stages of the universe, it is today confirmed beyond any doubt by means of several independent observations that the universe has again started to expand  at an accelerated rate approximately $6 \times 10^9$ years ago. A satisfactory explanation of this current acceleration that takes place at energy scales $10^{-4}\, {\rm eV}$ where we supposedly know physics very well is lacking. The most successful cosmological model accommodating the observed pattern of  expansion of the universe so far is the six parameter base $\Lambda$CDM model that is simple and in reasonably good agreement with available high precision data \cite{WMAP11a,Planck13}.
Yet two well-known problems related with the $\Lambda$ assumption, namely the coincidence and fine-tuning problems, may be signaling that it should rather be regarded as a good approximation or a limiting case of a more general model in which the vacuum energy need not necessarily be realized in terms of a cosmological constant $\Lambda$ and may deviate from $\Lambda$ considerably 
in the far past and/or future \cite{Zeldovich,Weinberg89,Sahni00,Peebles03}.
This calls for a dynamical description of dark energy and/or a possible modification of general relativity. Some recently reported tensions between the $\Lambda$ assumption and high precision data could also be resolved in case of evolving dark energy (DE) and/or a modified theory of gravity \cite{Sahni14,BOSS14,Vazquez12,Akarsu15v}.

\medskip

\noindent It is of course more desirable if inflation, or evolution scenarios that can be alternative to inflation, and the late time acceleration of the universe can all be derived from first principles or from a unified theory of fundamental interactions of nature such as the string theories or M-theory. A common property of such unified theories is that their consistent constructions require the existence of extra spatial dimensions in addition to our familiar four space-time dimensions. It is generally argued that we live in a universe in which all but four of the space-time dimensions are compactified on an unobservable, non-singular internal space without boundary; thus leaving behind an observable $(3+1)$-dimensional external space-time. But it is also known that the accelerated expansion of our external space is difficult to maintain in a cosmological context with such an approach unless the dimensional reduction of the internal space is dynamical \cite{freund,dereli1,shafi}.
It should not be surprising that considerations of extra dimensions also led to generalizations of Einstein's gravity in a natural way 
\cite{fradkin1, fradkin2, Callan86, Gross87, Boulware85, Zwiebach85, zumino, deser2, dereli2}. In fact long before the string models, specific higher dimensional generalizations of the Einstein tensor were pointed out such that
they are covariantly constant and contain at most second derivatives of the metric components \cite{lovelock1,lovelock2}. These tensor concomitants follow by a variational principle from  the dimensionally continued Euler-Poincar\'{e} densities.
The standard Kaluza-Klein reduction of  gravitational actions that include dimensionally continued Euler-Poincar\'{e} forms exhibits a complicated pattern of non-minimal interactions in four dimensions \cite{dereli3, dereli4, dereli5}. Their various cosmological solutions have also been much studied \cite{madore, muller-hoissen, Ishihara86,lorentz-Petzold, deruelle, dereli6}. However, the relatively recent notion of dark energy invoked a renewed upsurge of interest in their cosmological solutions \cite{Calcagni05, Nojiri05, Cognola06, Andrew07, Leith07, Bamba07, Chingangbam08}.

\medskip

\noindent In this paper, for the sake of simplicity, we are going to neglect the dilaton and axion fields that appear in effective string field theories. On the other hand, a cosmological constant/vacuum energy is not expected to occur in the effective low-energy theories of superstrings, but its existence is not excluded and it typically predicts negative vacuum energy, say a negative cosmological constant. Accordingly we consider also a negative cosmological constant in our model since its presence will lead to  more general models without complicating the field equations much.      
We are particularly interested in the dynamics of our external space in a higher dimensional steady state universe that was characterized by two main assumptions in \cite{Akarsu13a, Akarsu13b}: (i) the higher dimensional universe has a constant volume as a whole but both the internal and external spaces are dynamical. (ii) The energy density is constant in the higher dimensional universe. In this paper, we consider a higher dimensional fluid to model the matter distribution in the universe that exerts pressures that are distinct along the external and internal dimensions in accordance with the first assumption. Introduction of a higher dimensional anisotropic fluid provides us with new degrees of freedom that would in return allow us to deal with the dimensional dichotomy of our higher dimensional steady state universe without over-determining the system of field equations. We will be relaxing the second assumption accordingly and instead study a more general case in which the equation of state parameters of the higher dimensional fluid along the external and internal dimensions are dynamical but the difference between them is kept constant.

\medskip

\noindent In section \eqref{Section:2}, we give the gravitational field equations derived by a variational principle from a truncated action that includes only up to second order Euler-Poincar\'{e} forms in the gravitational sector. The vanishing of the space-time torsion is 
handled by considering  Palatini-type variations using the method of Lagrange multipliers. In section \eqref{Section:3} the cosmological space-time geometry is specified through a spatially homogeneous and flat but not necessarily isotropic $(1+3+n)$-dimensional synchronous space-time metric in locally Cartesian coordinates.  We also write explicitly the stress-energy-momentum tensor of a higher dimensional fluid. The reduced field equations are found. 
In section \eqref{Section:4}, the constant volume condition is imposed and the field equations are further reduced. 
We also discuss an equation of state that respects the dimensional dichotomy of the spatial geometry.  
Then exact solutions are found and classified according to the ranges of our free parameters. Section \eqref{Section:5} is devoted to concluding remarks. 


\section{Variational Field Equations}
\label{Section:2}
\noindent Einstein's gravitational field equations in 4-dimensions
is a non-linear system of second order partial differential equations that can be derived
by a variational principle from the Einstein-Hilbert action that involves a Lagrangian density
linear in the curvature components. It is not equally well-known that in dimensions higher than four,
a gravitational Lagrangian density can be supplemented
by unique additional terms, besides the Einstein-Hilbert term and a possible cosmological constant, without destroying the desirable property that the variational field equations do not carry derivatives of the metric components of order greater than two. Such additional terms in the action are essentially topological in nature and are given by the integrals of dimensionally continued Euler-Poincar\'{e} forms over the space-time manifold $M$. The zeroth order Euler-Poincar\'{e} form is taken to be the volume form itself and yields the cosmological constant. The first order Euler-Poincar\'{e} form is linear in the curvature components and is equal to the Einstein-Hilbert Lagrangian density. In $D=2$ dimensions this is a closed form
whose integral over a 2-manifold $M$ with boundary leads to the famous Gauss-Bonnet theorem. In dimensions $D>2$ its variations lead to Einstein field equations. In a similar way, the second order Euler-Poincar\'{e} form that is  quadratic in the curvature components is a closed form in $D=4$ dimensions. Its integral over a 4-manifold $M$ gives the topological Euler number of $M$.
The second order Euler-Poincar\'{e} density when continued to lower dimensions ($D<4$) vanishes identically,
while when continued to higher dimensions ($D>4$) its variations lead to second order dynamical field equations. The higher dimensional modified gravitational action obtained by  integrating the second order Euler-Poincar\'{e} form over $M$
is sometimes called the Gauss-Bonnet gravity in current literature. It was further pointed out by Lovelock long time ago that the $n$-th order Euler-Poincar\'{e} form is a closed form in $D=2n$ dimensions and if used as part of a gravitational action in dimensions $D>2n$ contributes only in second order to the highly non-linear variational field equations.

\medskip

\noindent Thus  the most general gravitational Lagrangian density
in $D$-dimensions that  yields second order variational field equations can be written as a linear combination of all possible dimensionally continued Euler-Poincar\'{e} densities:
\begin{equation}
L_{g} = \sum_{n=0}^{[\frac{D-1}{2}]} \Lambda_{n} L^{(n)}
\end{equation}
 where we introduced (dimension-full) coupling constants  $\Lambda_{n}$ and the  Euler-Poincar\'{e} forms are given by
 \begin{equation}
 L^{(n)} = \frac{1}{2^n} R_{a_1 b_1}  \wedge \dots \wedge R_{a_n b_n} \wedge *e^{a_1 b_1 \dots a_n b_n} .
\end{equation}
$R^{a}_{\;\;b} = d\omega^{a}_{\;\;b} + \omega^{a}_{\;\;c} \wedge \omega^{c}_{\;\;b}$ are the Riemann curvature 2-forms of (torsion-free)
Levi-Civita connection 1-forms $\{  \omega^{a}_{\;\;b} \}$. They are uniquely determined by a Lorentzian signatured metric
 tensor $g = \eta_{ab} e^{a} \otimes e^{b}$ where $\eta_{ab} = diag(-+++)$, given in terms of an orthonormal co-frame $\{e^a\}$, through the Cartan structure equations $de^a + \omega^{a}_{\;\;b} \wedge e^b = 0$. We further introduced the notation
 $ e^{ab...c} = e^{a} \wedge e^{b} \wedge ... \wedge e^{c}$
and the Hodge $*$-map that fixes the space-time orientation with the volume $D$-form $*1 = e^{012 \dots D-1}.$

\medskip

\noindent Here we vary the total action (in $D\geq 5$ dimensions)
\begin{equation}\label{totalaction}
I[e,\omega,\mu,\dots] = \int_{M} \left ( L_{g}  + L_{c} + \kappa L_{m} \right )
\end{equation}
where we take a truncated gravitational Lagrangian density given by
\begin{equation}
L_{g} = \lambda_{0} *1 + \frac{1}{2}R_{ab} \wedge *e^{ab} + \frac{\lambda_2}{4} R_{ab} \wedge R_{cd} \wedge *e^{abcd} ,
\end{equation}
where $\lambda_0$ stands as a higher dimensional negative cosmological constant. The constraint Lagrangian
\begin{equation}
L_{c} = \left ( de^a + \omega^{a}_{\;\;b} \wedge e^b \right ) \wedge \mu_{a}
\end{equation}
imposes the zero-torsion constraint on the connection through the variations of the Lagrange multiplier $(D-2)$-forms $\mu_a$.
$\kappa>0$ denotes the Newton's  constant that is scaled appropriately in $D$-dimensions. Then the co-frame variations of the action \eqref{totalaction} give
\begin{equation}\label{coframevary}
\lambda_0 *e_a + \frac{1}{2} R^{bc} \wedge *e_{abc} + \frac{\lambda_2}{4} R^{bc} \wedge R^{df} \wedge *e_{abcdf} +D\mu_a +\kappa \tau_{a}(m) = 0,
\end{equation}
while the connection variations of \eqref{totalaction} give
\begin{equation}\label{connectionvary}
*e_{abc} \wedge T^{c} + \frac{\lambda_2}{2} *e_{abcdf} \wedge R^{cd} \wedge T^{f} + \kappa \Sigma_{ab}(m) = e_{a} \wedge \mu_{b} - e_{b} \wedge \mu_{a}.
\end{equation}
In these expressions we introduced through  the variations of the matter Lagrangian, the stress-energy-momentum tensor $T_{ab}(m)$ and the angular momentum tensor $S_{ab,c}(m)$ to be read from
\begin{equation}
\left ( \frac{\delta L_{m}}{\delta e^a}\right ) = \tau_{a}(m) \equiv T_{ab}(m) *e^b
\end{equation}
and
\begin{equation}
\left ( \frac{\delta L_{m}}{\delta \omega^{ab}} \right ) = \Sigma_{ab}(m) \equiv S_{ab,c}(m) *e^c,
\end{equation}
respectively. The variations of the action with respect to the Lagrange multipliers impose the zero-torsion constraint $T^a = 0$ and the field equations \eqref{coframevary} and \eqref{connectionvary} above are to be solved subject to this constraint. Then the connection-variation equations simplify and an algebraic solution for the Lagrange multiplier forms can be found as follows:
\begin{equation}
\mu_a = \kappa \left ( \iota_{X_b}\Sigma_{a}^{\;\;b} - \frac{1}{2} e_{a} \wedge \iota_{X_c}\iota_{X_b}\Sigma^{bc} \right ).
\end{equation}
We use interior products such that $\iota_{X_a}(e^b) = \delta^{b}_{\;\;a}$ with respect to orthonormal frame vectors $\{ X_a \}.$
Then the Lagrange multipliers are substituted into the remaining field equations and we arrive at the final form of our modified Einstein field equations
\begin{eqnarray}\label{modEinstein}
\lambda_0 *e_a &+& \frac{1}{2} R^{bc} \wedge *e_{abc} + \frac{\lambda_2}{4} R^{bc} \wedge R^{df} \wedge *e_{abcdf} \nonumber \\ &=& -\kappa \tau_{a}(m)
+ \kappa D(\iota_{X_b}\Sigma_{a}^{\;\;b}) +  \frac{\kappa}{2} e_{a} \wedge D( \iota_{X_b}\iota_{X_c}\Sigma^{bc}).
\end{eqnarray}
As a last remark we note that the integrability of the Einstein field equations \eqref{modEinstein} yields the conservation law of matter:
\begin{equation}
D\tau_{a}(m) = R_{ab} \wedge \iota_{X_c}\Sigma^{cb}(m).
\end{equation}

\section{Cosmological Ansatz}\label{Section:3}

\noindent The geometry of a cosmological space-time will be given by the metric
\begin{equation}
g = -dt^2 + R(t)^2  \sum_{i=1}^{3} ({dx^i})^2 + S(t)^2 \sum_{\alpha=1}^{n} ({dy^\alpha})^2
\end{equation}
where $\{x^i\}$ are the Cartesian coordinates of a flat(external) 3-space and
$\{y^\alpha\}$ are the coordinates of a (locally) flat, compact (internal) n-space. $R(t)$ and $S(t)$ are the corresponding
scale factors of cosmic time $t$. Let us choose the following orthonormal basis 1-forms
\begin{equation}
e^0 = dt, \quad e^i = R(t) dx^i, \quad e^\alpha = S(t)dy^\alpha ,
\end{equation}
so that the non-zero connection 1-forms are determined to be
\begin{equation}
\omega^{0}_{\;\;i} = \frac{\dot{R}}{R} e^i ,\quad \omega^{0}_{\;\;\alpha} = \frac{\dot{S}}{S} e^{\alpha},
\end{equation}
where $^{.} = \frac{{\rm d}}{{\rm d} t}$ denotes time derivative. Therefore the curvature 2-forms are given by
\begin{eqnarray}
R^{0}_{\;\;i} &=& \left ( \frac{\ddot{R}}{R} + \frac{\dot{R}^2}{R^2} \right ) e^{0} \wedge e^{i}, \quad  \quad  R^{0}_{\;\;\alpha} = \left ( \frac{\ddot{S}}{S} +\frac{\dot{S}^2}{S^2}\right )e^{0} \wedge e^{\alpha}, \nonumber \\ R^{i}_{\;\;j} &=& \frac{\dot{R}^2}{R^2} e^{i} \wedge e^{j}, \quad  \quad R^{\alpha}_{\;\;\beta} = \frac{\dot{S}^2}{S^2} e^{\alpha} \wedge e^{\beta},\quad \quad R^{i}_{\;\;\alpha} = \frac{\dot{R}}{R}\frac{\dot{S}}{S} e^{i} \wedge e^{\alpha}.
\end{eqnarray}

\medskip

\noindent We will model the matter distribution  in a $D$-dimensional space-time  by a homogeneous but anisotropic fluid with no rotation. Thus we take $\Sigma_{ab}(m)=0$ and the stress-energy-momentum tensor of the fluid will be specified by
\begin{equation}
T=T_{ab}e^{a} \otimes e^{b} = \rho(t) dt^2 + p_{ext}(t)\sum_{i=1}^{3}(dx^i)^2  + p_{int}(t) \sum_{\alpha=1}^{n}(dy^\alpha)^2
\end{equation}
where we assume that the mass density $\rho >0$,  but the pressures $p_{ext}(t)$ and $p_{int}(t)$ are arbitrary except for the conservation law they should satisfy:
\begin{equation}
\dot{\rho} + 3\frac{\dot{R}}{R} ( \rho + p_{ext}) + n \frac{\dot{S}}{S} ( \rho + p_{int}) =0.
\end{equation}

\medskip

\noindent At this point, we skip calculational details and give the independent equations of motion we obtained as follows:
\begin{eqnarray}\label{19}
\lambda_0 &+&  3 \left\lbrace \frac{\dot{R}^2}{R^2} +  3 n \frac{\dot{R}}{R}\frac{\dot{S}}{S} + \frac{n(n-1)}{2} \frac{\dot{S}^2}{S^2}  \right\rbrace
 \nonumber \\  &+& \lambda_2 \left \{ 9n(n-1) \frac{\dot{R}^2}{R^2} \frac{\dot{S}^2}{S^2}  \right . \nonumber \\ &+& \left . 6n \frac{\dot{R}^3}{R^3}  \frac{\dot{S}}{S}
+ 3 n(n-1)(n-2) \frac{\dot{R}}{R} \frac{\dot{S^3}}{S^3} \right . \nonumber \\ &+& \left . \frac{n(n-1)(n-2)(n-3)}{4} \frac{\dot{S}^4}{S^4} \right \}  = \kappa \rho ,
\end{eqnarray}
\begin{eqnarray}
\label{20}
\lambda_0 &+&  \left \{ 2\frac{\ddot{R}}{R} + \frac{\dot{R}^2}{R^2} + n \frac{\ddot{S}}{S}  + 2 n \frac{\dot{R}}{R}\frac{\dot{S}}{S} +
\frac{ n(n-1)}{2} \frac{\dot{S}^2}{S^2} \right \} \nonumber \\
 &+& \lambda_2 \left \{ \frac{\ddot{R}}{R}  \left( 2n(n-1) \frac{\dot{S}^2}{S^2} + 4n \frac{\dot{R}}{R}\frac{\dot{S}}{S} \right) \right . \nonumber \\ &+& \left . \frac{\ddot{S}}{S} \left( 2n \frac{\dot{R}^2}{R^2} +  4n(n-1) \frac{\dot{R}}{R}\frac{\dot{S}}{S}+  n(n-1)(n-2) \frac{\dot{S}^2}{S^2} \right) \right. \nonumber \\
 &{}& \left. +3n(n-1) \frac{\dot{R}^2}{R^2} \frac{\dot{S}^2}{S^2} + 2n(n-1)(n-2) \frac{\dot{R}}{R} \frac{\dot{S^3}}{S^3} \right . \nonumber \\ &+& \left . \frac{n(n-1)(n-2)(n-3)}{4} \frac{\dot{S}^4}{S^4} \right \} = -\kappa p_{ext},
\end{eqnarray}
\begin{eqnarray}
\label{21}
\lambda_0 &+& \left \{ 3\frac{\ddot{R}}{R}+(n-1)\frac{\ddot{S}}{S} + 3 (n-1) \frac{\dot{R}}{R} \frac{\dot{S}}{S} + 3 \frac{\dot{R}^2}{R^2}  + \frac{(n-1)(n-2)}{2} \frac{\dot{S}^2}{S^2} \right \} \nonumber \\
 &+& \lambda_2 \left \{ \frac{\ddot{R}}{R}\left(6(\frac{\dot{R}}{R})^2+ 12(n-1) \frac{\dot{R}}{R}\frac{\dot{S}}{S} +3(n-1)(n-2) \frac{\dot{S}^2}{S^2} \right) \right. \nonumber \\
&+& \left . \frac{\ddot{S}}{S} \left ( 6(n-1)\frac{\dot{R}^2}{R^2} + 6(n-1)(n-2)\frac{\dot{R}}{R} \frac{\dot{S}}{S}+(n-1)(n-2)(n-3) \frac{\dot{S}^2}{S^2} \right ) \right . \nonumber \\ &{}& \left . + 9(n-1)(n-2) \frac{\dot{R}^2}{R^2} \frac{\dot{S}^2}{S^2}
+ 6(n-1) \frac{\dot{R}^3}{R^3} \frac{\dot{S}}{S} \right . \nonumber \\ &{}& \left . + 3(n-1)(n-2) (n-3) \frac{\dot{R}}{R} \frac{\dot{S^3}}{S^3} \right . \nonumber \\ &+& \left . \frac{ (n-1)(n-2)(n-3)(n-4)}{4} \frac{\dot{S}^4}{S^4} \right \}  = -\kappa p_{int}.
\end{eqnarray}
The coupled equations \eqref{19}, \eqref{20} and \eqref{21}  are to be satisfied by five functions $R$, $S$, $\rho$, $p_{ext}$ and $p_{int}$ and therefore the system is not fully determined as it is. Two extra constraints must be provided by some extra assumptions in order to get fully determined solutions.


\section{Exact Solutions with Constant Volume}\label{Section:4}

\noindent We are particularly interested in higher dimensional steady state cosmologies characterized by the following two properties: (i) the higher dimensional universe has a constant volume as a whole but the internal and external spaces are dynamical. (ii) The mass density is constant in the higher dimensional universe. These are strong assumptions that had been motivated and discussed in the context of similar cosmological models in our previous works \cite{Akarsu13a,Akarsu13b}. We will further add comments towards their justification in the concluding section \ref{Section:5}. Here it should be noted, however, that we allow the energy density to be dynamical in a certain way, so that a constant mass density can be studied as a special case. Therefore, firstly we assume that the $(3+n)$-dimensional \textit{volume scale factor} of the universe is constant:
\begin{equation}\label{constvol}
R^{3}S^{n}=V_{0}.
\end{equation}
It will be convenient to introduce the Hubble functions of external and internal spaces given by $H =  \frac{\dot{R}}{R}$ and $H^{\prime}=\frac{\dot{S}}{S} $, respectively. Then it follows from \eqref{constvol} that $3H =- nH^{\prime} $, assuring the dynamical contraction (hence reduction) of the internal space ($H^{\prime}<0$) for an expanding external space ($H>0$). The field equations  in terms of $H$ read:
\begin{equation}\label{rhoH}
\lambda_0 - \frac{3n+9}{2n} H^2 +  3 \lambda_2 b_n H^4 = \kappa \rho ,
 \end{equation}
 \begin{equation}\label{pextH}
 \lambda_0 -\dot{H} + \frac{3n+9}{2n}H^2 +\lambda_2 H^2 \left ( a_n\dot{H} - b_n H^2 \right) =-\kappa p_{ext} ,
\end{equation}
\begin{equation}\label{pintH}
 \lambda_0  + \frac{3}{n} \dot{H} + \frac{3n+9}{2n} H^2 + \lambda_2 H^2 \left (  (a_n-4b_n) \dot{H} - b_n H^2  \right) =-\kappa p_{int}
\end{equation}
where we set\footnote{Note that  $a_n$ and $b_n$ are both negative  for $n=1$ and both positive for all $n\geq 2$. For the sake of simplicity we will keep the second case in what follows.}
\begin{equation}
 a_n =\frac{9\left(n^2+3 n-6\right)}{n^2}, \quad
  b_n =  \frac{3(n+3)(n^2+ 15n-18)}{4n^3}.
\end{equation}
Furthermore our conservation law of matter reduces to
\begin{equation}\label{conserveqn}
\dot{\rho} + 3H \left ( p_{ext}-p_{int} \right ) = 0 .
\end{equation}
Secondly we assume a simple equation of state
\begin{equation}\label{EoS}
\frac{ p_{ext}-p_{int}  }{\rho} = \Delta w
\end{equation}
where $\Delta w$ is a constant that we propose to call dimensional dichotomy parameter. {It follows that $\rho \propto R^{-3\Delta w}$ in general. A higher dimensional isotropic distribution of matter with constant mass density follows in the case $\Delta w=0$. If there is a dimensional dichotomy i.e. $\Delta w\neq 0$, on the other hand, then, as the external space expands, the mass density decreases if $p_{ext}>p_{int}$ and increases if $p_{ext}<p_{int}$.}

\medskip

\noindent In order to solve this system we proceed in the following way. We differentiate both sides of \eqref{rhoH} and on the RHS use the
conservation relation \eqref{conserveqn} together with our equation of state \eqref{EoS} to replace $\dot{\rho}$.
Then we use the first equation \eqref{conserveqn} again to eliminate $\rho$ itself, arriving at a first order master equation satisfied by $H$.
We will solve this master equation and classify its solutions according to the signs and magnitudes of the cosmological coupling constants $\lambda_0$ and $\lambda_2$. Once an expression for $H$ is fixed, the external and internal pressure functions can be found by substituting this $H$ into  the equations \eqref{pextH} and \eqref{pintH}, respectively. Finally we integrate $H$ and determine the scale factors $R$ and $S$, thus completely specifying the cosmology.

 \medskip

\noindent From now on we will be looking at the master equation for $H$ that reads
\begin{eqnarray}
\left [ \left (\frac{3n+9}{2n}\right ) H^2 \right . &-& \left . \frac{n(n+3)}{2(n^2+15n-18)}\frac{1}{\lambda_2}\right ] \left ( \frac{3n+9}{2n} \right )\dot{H}
\nonumber \\ &=&
-\frac{3\Delta w}{4} \left [\left ( \frac{3n+9}{2n} \right )^2 H^4 - \frac{n(n+3)}{(n^2+15n-18)} \frac{1}{\lambda_2} \left ( \frac{3n+9}{2n} \right ) H^2 \right . \nonumber \\  &+& \left . \frac{n(n+3)}{(n^2+15n-18)} \frac{\lambda_0}{\lambda_2}\right ].
\end{eqnarray}
We
re-write it as a simple differential equation
\begin{equation}
\dot{x} = \omega \left [ (x^2 - \lambda) - \frac{Q}{(x^2 - \lambda)} \right ]
\end{equation}
where we defined
\begin{equation}
x \equiv \sqrt{\frac{3n+9}{2n}} H, \quad \omega \equiv -\sqrt{\frac{2n}{3n+9}} \frac{3\Delta w}{4}
 \end{equation}
with two free parameters
\begin{equation}
\lambda \equiv \frac{n(n+3)}{(n^2+15n-18)} \frac{1}{2\lambda_2} , \quad Q \equiv  \lambda^{2} - 2 \lambda_0 \lambda.
\end{equation}
It is worth to note that the field equation \eqref{rhoH} with the above re-parametrization gives for the energy density
\begin{equation}
\kappa \rho = \frac{1}{2\lambda} \left [ (x^2 - \lambda)^2 - Q \right ].
\end{equation}

\bigskip

\noindent We first consider briefly the case of an isotropic distribution of matter. Suppose $\lambda \neq 0$. We take $\omega=0$ and
the master equation becomes $\dot{x}=0$. This leads to an exponentially expanding external scale factor $R(t)=R(0)e^{H_0 t}$ and an exponentially
contracting internal scale factor $S(t)=S(0)e^{-\frac{3}{n}H_0 t}.$ For isotropic solutions with $\lambda \neq 0$, the energy density may be positive definite or negative definite or zero depending on the roots of the algebraic equation
$$
\lambda_0 - \frac{3n+9}{2n}H_{0}^{2} + \frac{1}{2\lambda} \left(\frac{3n+9}{2n}\right)^2 H_{0}^{4} = 0. 
 $$

\medskip

\noindent Next we consider cases  with $\omega \neq 0$ and $\lambda \neq 0$:

(1) Suppose $Q \neq 0$. The generic solution is implicitly given by
\begin{equation}
\left | \frac{\sqrt{\lambda+\sqrt{Q}}-x}{ \sqrt{\lambda+\sqrt{Q}}+x } \right |^{\frac{1}{2\sqrt{\lambda+\sqrt{Q}}}}  \left | \frac{\sqrt{\lambda -\sqrt{Q}} -x }{\sqrt{\lambda -\sqrt{Q}}+x} \right |^{\frac{1}{2\sqrt{\lambda-\sqrt{Q}}}}  = Ce^{2\omega t}
\end{equation}
where $C$ is an integration constant. We cannot further develop such solutions and obtain explicit expression for $R(t)$ and $S(t)$ in general.

(2) Suppose $Q=0$, that is $\lambda = 2\lambda_0$. Now the generic solution reads
\begin{equation}
\left | \frac{\sqrt{\lambda}-x}{ \sqrt{\lambda}+x } \right |^{\frac{1}{\sqrt{\lambda}}}= Ce^{2\omega t},
\end{equation}
and can be  inverted to write down explicit expressions.
For $\lambda >0$ we have
\begin{equation}
x(t) = \left \{ \begin{array}{l} -\sqrt{\lambda} \coth(\sqrt{\lambda}\omega t), \quad  x^2 > \lambda \\
-\sqrt{\lambda} \tanh(\sqrt{\lambda}\omega t), \quad  x^2 < \lambda \end{array} . \right .
\end{equation}
These expressions can be integrated for the scale factor of the external space so that
\begin{equation}
\label{EHEPsinh}
R(t) = \left \{ \begin{array}{l}  |\sinh(\sqrt{\lambda}\omega t)|^{4/(3\Delta w)}, \quad  x^2 > \lambda \\
 |\cosh(\sqrt{\lambda}\omega t)|^{4/(3\Delta w)}, \quad  x^2 < \lambda \end{array} . \right .
\end{equation}
The corresponding mass density function turns out to be positive definite:
\begin{equation}
\label{rhoEHEPsinh}
\kappa \rho(t) = \left \{ \begin{array}{l}  \frac{\lambda}{2} \sinh^{-4}(\sqrt{\lambda}\omega t), \quad  x^2 > \lambda \\
 \frac{\lambda}{2}\cosh^{-4}(\sqrt{\lambda}\omega t) , \quad  x^2 < \lambda \end{array} . \right .
\end{equation}

\medskip

For $\lambda <0$ we have 
\begin{equation}
x(t)=-\sqrt{|\lambda|} \cot( \sqrt{|\lambda|}\omega t).
\end{equation}
Again we integrate the above expression and find that
\begin{equation}
R(t) =  |\sin(\sqrt{|\lambda|}\omega t)|^{4/(3\Delta w)}.
\end{equation}
The corresponding mass density function for this solution is negative definite:
\begin{equation}
\kappa \rho(t) = 
-\frac{|\lambda|}{2}\sin^{-4}(\sqrt{\lambda}\omega t).
 \end{equation}

\medskip

\noindent We also wish to point out some special classes of solutions which cannot be obtained from above as limiting cases.

(A) First consider the case of Einstein-Hilbert action with a cosmological constant where the Euler-Poincar\'{e} term is absent. We set $\lambda_2=0$  accordingly but still keep $\omega \neq 0$.
Then the master equation reads 
\begin{equation}
\dot{x} = 2\omega \left ( x^2-\lambda_0\right )
\end{equation}

For $\lambda_0 >0$ the solutions are given by
\begin{equation}
x(t) = \left \{ \begin{array}{l} -\sqrt{\lambda_0} \coth(2\sqrt{\lambda_0}\omega t), \quad  x^2 > \lambda_0 \\
-\sqrt{\lambda_0} \tanh(2 \sqrt{\lambda_0}\omega t), \quad  x^2 < \lambda_0 \end{array} . \right .
\end{equation}
The external scale factor can be calculated as
\begin{equation}
\label{EHsinh}
R(t) = \left \{ \begin{array}{l}  |\cosh(2 \sqrt{\lambda_0}\omega t)|^{2/(3\Delta w)}, \quad  x^2 > \lambda_0 \\
 |\sinh(2\sqrt{\lambda_0}\omega t)|^{2/(3\Delta w)}, \quad  x^2 < \lambda_0 \end{array} , \right .
\end{equation}
which give a positive definite the energy density for $0<\lambda_0<x^2$
\begin{equation}
\kappa \rho(t) = \frac{\lambda_0}{\cosh^{2}(2\sqrt{\lambda_0}\omega t)},
\end{equation}
and a negative definite energy density for $x^2<\lambda_0$
\begin{equation}
\label{rhoEHsinh}
\kappa \rho(t) = -\frac{\lambda_0}{\sinh^{2}(2\sqrt{\lambda_0}\omega t)}.
\end{equation}

For $\lambda_0 <0$ we have
\begin{equation}
x(t) =  \sqrt{|\lambda_0|} \tan (2\sqrt{|\lambda_0|}\omega t)
\end{equation}
and
\begin{equation}
R(t)= |\cos(2\sqrt{|\lambda_0|}\omega t)|^{2/(3\Delta w)}.
\end{equation}
In this case, the corresponding energy density is always negative definite:
\begin{equation}
\kappa \rho = -\frac{|\lambda_0|}{\cos^{2}(2\sqrt{|\lambda_0|}\omega t)}.
\end{equation}

For the particular case $\lambda_0=0$ we have the following external scale factor
\begin{equation}
R(t)\propto t^{2/(3 \Delta w)},
\end{equation}
which leads to the following negative definite mass density
\begin{equation}
\kappa \rho= -\frac{1}{4 \omega^2 t^2}.
\end{equation}
{It is easy to see that, in this particular case $\lambda_0 =  0$, the external space exhibits accelerating expansion provided that $0<\Delta w< \frac{2}{3}$. However, we note that it suffers from the negativity of mass density.}

\medskip

(B) Suppose the gravitational action consists of just the second order Euler-Poincar\'{e} term on its own. This theory is sometimes called the Gauss-Bonnet gravity. Let us consider the cases where $\omega \neq 0$. Then the master equation becomes
\begin{equation}
\dot{x}=\omega x^2,
\end{equation}
which gives the following a power-law expansion
\begin{equation}
R(t) \propto t^{4/(3{\Delta w})},
\end{equation}
and the following mass density
\begin{equation}
\rho=\frac{1}{2\lambda\omega^4}\frac{1}{t^4}.
\end{equation}
{It is easy to see that in this solution the external space exhibits accelerating expansion if $0<\Delta w< \frac{3}{4}$. Additionally the mass density takes positive values properly provided that $\lambda$ is positive, i.e., $\lambda_2$ is positive in line with string theories.}

\medskip

\noindent Among the explicit solutions we presented above, the case for which the external space expands as $R \propto |\sinh(\sqrt{\lambda}\omega t)|^{4/(3\Delta w)}$ given in (\ref{EHEPsinh}) is the most promising solution from the cosmological point of view. We note that in this solution the external space behaves as $R\sim t^{4/(3\Delta w)}$ for $t\sim 0$ and $R \sim e^{\frac{4\sqrt{\lambda} \omega}{3\Delta w}\,t}$ for $t\gg \frac{1}{\sqrt{\lambda}\omega}$. According to this, provided that $\Delta w >3/4 $, the external space evolves from a decelerating expansion phase to the de Sitter expansion as the cosmic time $t$ runs. Moreover, if we choose $\Delta w=2$ then the external space exhibits exactly the same behavior as the standard $\Lambda$CDM model \cite{Sahni00}, namely, $R_{\Lambda{\rm CDM}}\propto \sinh^{\frac{2}{3}}\left(\sqrt{\frac{3\Lambda}{4}}\,t\right)$, where $\Lambda$ is the conventional positive cosmological constant. We note that similar results could be reached in the solution (\ref{EHsinh}) where the second order Euler-Poincar\'{e} term was omitted. However, we notice an important difference between these two solutions when we consider the corresponding matter densities. The solution (\ref{EHsinh}) where the second order Euler-Poincar\'{e} term was omitted suffers from negativity of mass density (\ref{rhoEHsinh}), while the solution (\ref{EHEPsinh}) where all the Einstein-Hilbert and the second order Euler-Poincar\'{e} terms are considered leads properly to positive definite mass density (\ref{rhoEHEPsinh}).

\section{Concluding Remarks}\label{Section:5}

We are interested in the idea of a higher dimensional steady state universe model because it combines in an interesting way the conventional ideas of a Big Bang cosmology with the steady state universe by going to higher dimensions. Namely, since the total $(3+n)$-dimensional volume of the universe is assumed to be constant, the universe in higher dimensions is eternal with no beginning and no end. But on the other hand the extra space dimensions are expected to contract dynamically to an unobservable size (dynamical reduction) while the three dimensional space would be expanding (external space), hopefully, in accordance with its evolutionary pattern that we observe today. The accelerated expansion of the 3-space in this picture would be driven by  mass that is being continuously transferred from the contracting $n$-dimensional internal space into the expanding $3$-dimensional external space. Hence the eternal amount of matter is fixed. It is  neither created nor exhausted but is redistributed between the external and internal spaces. 

In fact some other higher dimensional cosmological models whose solutions yield a constant higher dimensional volume exist in recent literature (See for instance \cite{Canfora13, Ohta14, Chirkov14a, Chirkov14b, Ivashchuk15}). Yet, the predicted behavior of the external space in these papers is not rich enough to explain the observed accelerated expansion, as these solutions exhibit either the power-law expansion as in Ref.\cite{freund} or de Sitter expansion as in Ref.\cite{dereli1}. In two recent papers, on the other hand, two of the present authors (O.A. and T.D.) considered  higher dimensional steady state models by imposing the constant volume condition from the start and then studied the dynamics implied by this assumption. In \cite{Akarsu13a}, the authors considered general relativity in $(1+3+n)$-dimensions and obtained various interesting new dynamics for the external space that yield a time varying deceleration parameter with oscillating cases included when flat/curved external and curved/flat internal spaces are considered. In \cite{Akarsu13b}, they extended the study by considering dilaton gravity, that mimics the low energy effective string theory action in a very simple manner when the dilaton coupling paramater $w$ is fixed to unity. It is found that among all possible values in the parameter space of the dilaton coupling parameter and the number of the internal dimensions, the case that corresponds to the bosonic string theory ($w=1$ and $n= 22$) gives the best history for the effectively four dimensional universe as it fits very well the standard history of the universe. Moreover, the case $w=1$ with $n = 22$ not only predicts the time and redshift values at which the accelerated expansion has started in accordance with the observations, but also predicts the $^4$He mass fraction in agreement with the mostly used observations for the Big Bang Nucleosynthesis. These promising results were obtained  using gravitational field equations motivated by effective string field theory actions at tree level. In this paper within the context of higher dimensional steady state universe idea we find it natural to investigate a modified gravity theory motivated by loop effects in superstring theories as a quadratic correction at the low energy limit. We presented various exact solutions by assuming a higher dimensional anisotropic fluid satisfying an equation of state with dimensional dichotomy in addition to the $(3+n)$-dimensional constant volume assumption. We showed that the external space in our model can exhibit exactly the same kinematics with the standard $\Lambda$CDM model even if we omit the second order Euler-Poincar\'{e} term. However the solution where the second order Euler-Poincar\'{e} term was omitted suffers from negativity of mass density, while the solution where all the Einstein-Hilbert and the second order Euler-Poincar\'{e} terms are considered can maintain the positivity of the mass density properly. Hence consideration of the higher order Euler-Poincar\'{e} terms could be crucial for the construction of promising higher dimensional steady state universe models.

\section*{Acknowledgments}
 \"{O}.A. acknowledges the support by T\"{U}B{\.I}TAK Research Fellowship for Post-Doctoral Researchers (2218). N.O. is also supported in part  by T\"{U}B{\.I}TAK with a Ph.D.Scholarship. We all appreciate the support from Ko\c{c} University.

\end{document}